# Demonstration of Broadband Non-Resonant Time-Crystal Amplification in Microwaves

Thomas R. Jones[1†], Ludmila J. Prokopeva[1†], Alexander V. Kildishev[1], Mordechai Segev[2], and Dimitrios Peroulis[1*]
[1] Elmore Family School of Electrical and Computer Engineering, Purdue University, West Lafayette, IN, USA
[2] Department of Physics, Technion-Israel Institute of Technology, Haifa, Israel
† equal contribution
[*] corresponding author: dperouli@purdue.edu

**Abstract:** We report an optically modulated experimental realization of a photonic time crystal (PTC) in the microwave regime, demonstrating for the first time that the PTC exponential growth can overcome losses and finite-size constraints of a practical spatio-temporal system and yield stable positive terminal gain over a continuous broadband frequency range. The developed experimental platform is a purely time-modulated capacitor (TMC) microwave circuit based on a microstrip transmission line, in which synchronized optical modulation of reverse-biased photodiodes generates strong (94.5 %) temporal modulation of the effective capacitance at 200 MHz. Broadband amplification consistent with a momentum band gap (MBG), a defining signature of photonic time-crystal physics, is observed, with a peak gain of 3.8 dB over a 65 MHz bandwidth. In addition, a narrow parametric resonance appears at the center of the band gap, reaching 4.8 dB. This sharp peak is associated with the spatial inhomogeneities of the lumped-element realization, while the corresponding homogeneous distributed system retains the Floquet-mode structure of a photonic time crystal. We show that finite microwave TMC implementations inherit the defining physics of PTCs, including phase-invariant non-resonant amplification and slow-light behavior inside the momentum band gap, while finite-size and loss mechanisms transform the ideal semicircular PTC gain profile into a continuous asymmetric non-Lorentzian gain band characterized by a Pearson type IV distribution.

---

## Introduction

Photonic time crystals (PTCs), electromagnetic media whose constitutive parameters are periodically modulated in time, have emerged as a fundamentally new platform for wave manipulation beyond the constraints of stationary materials [1–5]. Their origin can be traced to earlier studies of temporal interfaces and spacetime-varying media, where abrupt temporal modulation was shown to produce temporal reflection and temporal refraction, the temporal counterparts of conventional spatial scattering processes [6–8]. These developments established a new research direction at the intersection of photonics, metamaterials, Floquet physics, and strongly time-modulated wave systems, extending beyond the weakly perturbative regime of conventional nonlinear optics [2].

PTC concepts rapidly triggered extensive theoretical work on the unusual electrodynamics enabled by temporally periodic media [3–6,9–11]. These studies predicted momentum band gaps and non-resonant amplification [1,2], topological temporal edge states [2], temporal localization and disorder-induced amplification [12], spacetime crystals and nonreciprocal wave transport [6,9], temporal coatings and inverse-prism concepts [9,10], coherent absorption and amplification [13], and photonic time-crystal lasing [1]. The theoretical foundations of PTC electrodynamics continue to evolve, including generalized Kramers–Kronig relations for temporally dispersive media [14], Floquet scattering-matrix formalisms [15], homogenization theories [11,16], microscopic conservation laws and temporal boundary conditions [8], and, most recently, the interpretation of transport quantities in PTCs [17].

In parallel, experiments began to validate key predictions of time-varying photonics. Dynamically switched microwave and photonic metamaterials demonstrated temporal refraction/reflection, broadband frequency translation, and related temporal scattering phenomena [18,19]. Metasurface-based and temporally modulated microwave platforms subsequently demonstrated experimentally realizable photonic time crystals, momentum-gap amplification, and temporal topological states [20,21], while driven metamaterial systems exhibited photonic analogues of continuous time-crystalline behavior [22]. These experiments established temporally modulated photonic media as an experimentally accessible platform for Floquet wave physics.

Despite this rapid progress, whether the exponential growth associated with momentum band gaps can produce broadband and stable positive terminal amplification in practical photonic time-crystal devices has remained unresolved. In ideal infinite PTCs, modes residing inside the momentum band gap undergo non-resonant exponential amplification by continuously extracting energy from the temporal modulation [1,2]. Experimental studies reported signatures consistent with momentum-gap amplification and related Floquet instabilities [20,21], while ultrafast optical platforms demonstrated coherent amplification and coherent perfect absorption [13]. However, it remains unclear whether this growth can survive finite-size effects, dissipation, impedance mismatch, and modulation nonidealities sufficiently to produce broadband and stable positive gain in a finite propagating device. Prior demonstrations of photonic time-crystal amplification reported either internal modal amplification within momentum band gaps [20,21] or phase-sensitive interferometric gain [13], but did not achieve continuous-bandwidth positive transmission gain ($|S_{12}|>1$) between the input and output ports of a finite propagating system. Consequently, broadband terminal amplification arising directly from photonic time-crystal Floquet dynamics has remained experimentally unresolved until now.

## Theoretical Description

To connect the ideal photonic time crystal to the experimental microwave implementation, we first analyze a spatially uniform medium whose permittivity is periodically modulated in time (Periodic PTC, Fig. 1a). In this ideal PTC, temporal interfaces repeatedly generate time-reflected and time-refracted waves, which interfere constructively over finite intervals of momentum and open momentum band gaps (MBG). Inside the k-gap, the Bloch frequency ($\Omega$) becomes complex, and the positive imaginary branch corresponds to exponential field growth. The continuous lineshape of Im(Ω) can be derived analytically for the canonical square-wave PTC under the quarter-wave phase match condition ($p_1/p_2 = n_1/n_2$ – durations of states are proportional to their refractive indices),

$$\mathrm{Im}(\Omega)p = \cosh^{-1}\left[\frac{2}{1-(C_m^{sq})^2}\sin^2\left(\frac{kd}{2}\right)-1\right], \tag{1}$$

and is well approximated by a compact-support semicircular envelope with the center at $kd = \pi$, where gain maximum is

$$Im\frac{\Omega}{\omega_m} = \frac{1}{2\pi}\tanh^{-1}(C_m^{sq}). \tag{2}$$

For the experimentally relevant trapezoidal modulation, with finite rise and fall times (in our case, $t_{\mathrm{rise}}/p = t_{\mathrm{fall}}/p = 0.2$, since $t_{\mathrm{rise}} = 1$ns and $f_m = 200$ MHz), the same MBG can be matched by choosing a square modulation with depth adjusted according to Eq. (2). The trapezoidal waveform preserves the MBG location and instability strength but modifies the real part of the dispersion, producing asymmetric transport shoulders (also see measured velocity in Fig. 2c). For smooth time-domain TMC simulations, we use a single-harmonic modulation that primarily opens the fundamental MBG, with modulation depth chosen to reproduce the same peak Im(Ω),

$$Im\frac{\Omega}{\omega_m} = \frac{1}{8}\tanh^{-1}(C_m^{cos}). \tag{3}$$

In the finite PTC (Fig. 1b), where the modulation is applied for only a few periods ($N = 5$), the MBG remains visible as a continuous gain band in the transmission spectrum. Importantly, the MBG does not simply add gain on top of the temporal Fabry–Pérot oscillations, instead the ripples are suppressed inside the $k$-gap, and the spectrum follows the underlying analytical semicircle MBG envelope. This establishes the band-structure origin of the broadband gain in the ideal PTC.

We then map this ideal PTC response onto a time-modulated-capacitor (TMC) transmission-line model (Fig. cd). In the sub-wavelength limit, the lumped LC ladder approaches a distributed medium, with the capacitance modulation playing the role of a time-varying permittivity. Using experimentally realistic photodiode parameters, corresponding to a large effective capacitance contrast $C_m = (C_2 - C_1)/(C_2 + C_1) \approx 0.945$, the calculated TMC band structure retains the same MBG as the ideal PTC. Thus, before including losses or parasitics, the lossless TMC model predicts that sufficiently deep synchronized capacitance modulation should produce broadband amplification around $f_m/2$, with a gain bandwidth and center frequency set directly by the Floquet band structure. The finite TMC (Fig. 1d) converts this bulk instability into a measurable terminal gain. Cascading a small number of time-modulated unit cells (here 1-7) produces a continuous gain band whose envelope develops finite spectral tails. The broadband component grows with the number of cells and is insensitive to the absolute modulation phase, consistent with its origin as a band-structure effect. This phase invariance distinguishes the MBG amplification from ordinary resonant parametric gain.

The lumped TMC band structure (Fig. 1c) also predicts an additional narrow feature near the MBG center. Unlike the broadband MBG gain, this feature is absent in the homogeneous distributed limit and arises from the discrete circuit implementation, where spatial impedance discontinuities and temporal modulation combine to support a sharp mid-gap parametric resonance. Because this resonance is tied to the lumped realization rather than to the continuous MBG itself, it is expected to be more phase-sensitive and to show weaker scaling with the number of cells. This provides a natural interpretation of the experimentally observed response: a robust, phase-invariant broadband gain band produced by the PTC momentum band gap, with a superimposed narrow, phase-sensitive resonance at the gap center.

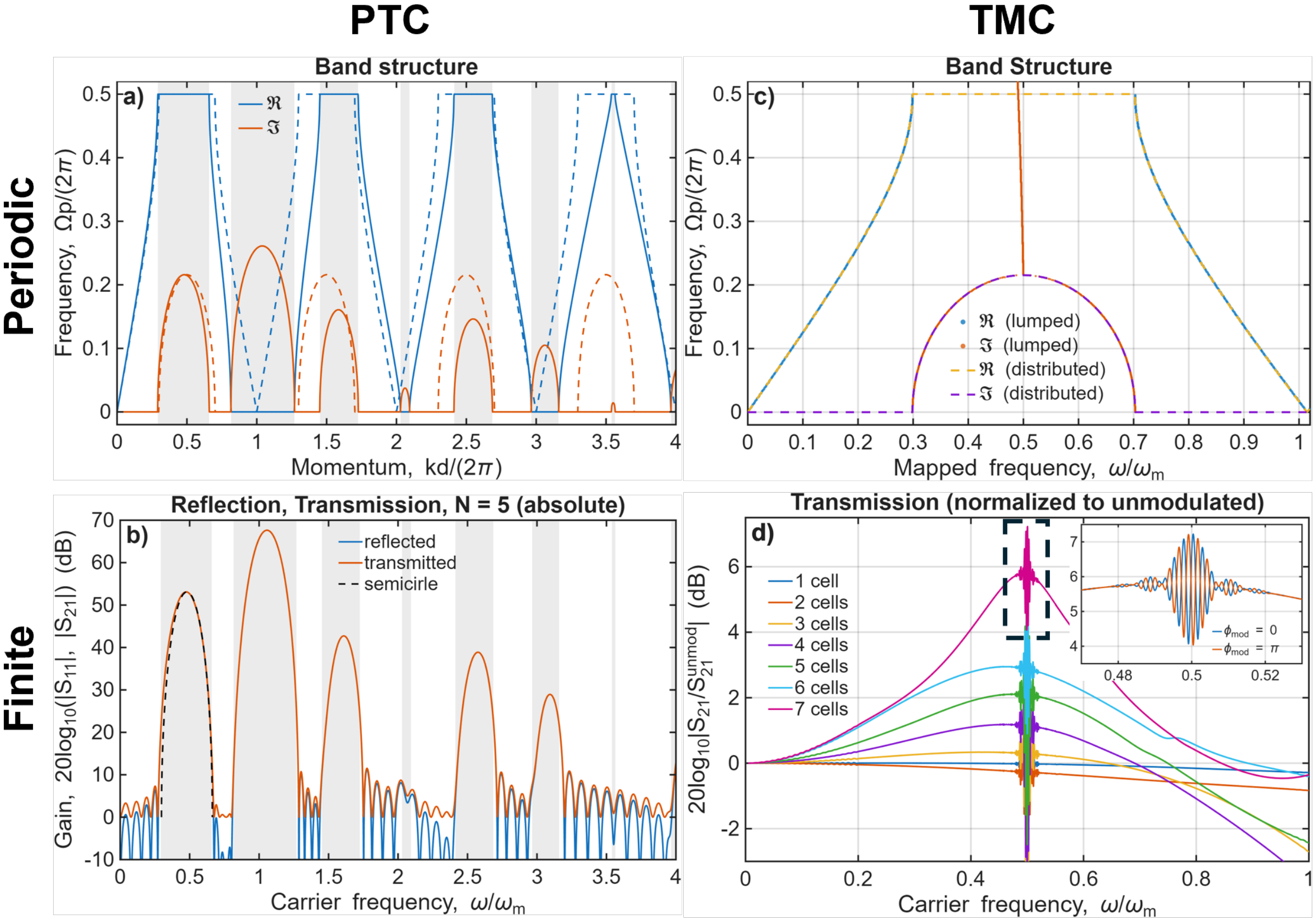

**Figure 1 | Theoretical description: From ideal photonic time crystals (PTC) to continuous terminal gain in finite Time-Modulated-Capacitor (TMC) circuits.** (a) Periodic PTC band structure of an ideal photonic time crystal for symmetric trapezoidal modulation with rise-time ratio $t_{rise}/p = 0.2$ (solid lines), compared with the quarter-wave-matched square-wave limit (dashed lines) with matched maximum of Im(Ω) of the fundamental harmonic. (b) Finite PTC: reflected and transmitted power for a medium modulated over $N = 5$ temporal periods. (c) Periodic TMC band structure: time-modulated LC ladder (series inductor L and shunt capacitor $C(t)$) under harmonic modulation $C(t)/C_a = 1 + C_m^{cos}\cos(\omega_m t)$ with matched $C_m^{cos} = \tanh(8 \cdot 0.216)$; for the lumped model, dispersion is mapped using subwavelength approximation $\frac{kd}{2} \sim \sin\left(\frac{kd}{2}\right)$. (d) Finite TMC: transmission through finite LC-ladder containing 1-7 unit cells, normalized to the corresponding unmodulated structure; Inset: transmission for 0- and $\pi$-shifted modulation phases. Notations: $f_m$ – modulation frequency ($\omega_m = 2\pi f_m$), $p = 1/f_m$ – modulation period, $d = n_{eff}^{-1} p c_0$ – modulation wavelength. Modeling parameters: modulation depth is $C_m = (C_2 - C_1)/(C_1 + C_2) = 0.945$ (equivalently, permittivity contrast $C_m = (\varepsilon_2 - \varepsilon_1)/(\varepsilon_1 + \varepsilon_2)$); the static LC-ladder stop band edge is placed at $f_{stop} = \frac{1}{\pi\sqrt{LC_a}} = 2f_m$ (well above $f_m/2$); loss and finite microstrip effects are neglected to isolate the underlying time-modulation physics.

## Experimental Demonstration

The experimentally realized TMC platform extends the idealized PTC/TMC model to a lossy microwave system containing finite-size circuit elements. The structures consist of periodically loaded microstrip transmission lines whose effective capacitance is synchronously modulated at $f_m = 200$ MHz using optically driven reverse-biased photodiodes (see Methods section for details). Unlike previously modeled PTC/TMC system, the experimental system contains finite propagation length, impedance mismatch, parasitic RC loading, and distributed microstrip dispersion. Nevertheless, the measured transmission spectra preserve the key signatures predicted by the theoretical description in the previous section.

Figure 2a compares measurements and full-wave circuit simulations for a five-cell TMC implementation. Two distinct amplification mechanisms are observed. First, a broadband continuous gain region emerges around $f_m/2$, consistent with the momentum-band-gap (MBG) instability predicted by the PTC dispersion. This broadband component scales with the number of unit cells and remains largely insensitive to the modulation phase (Fig. 2b). Second, a narrow peak appears at the exact gap center. Unlike the broadband MBG response, this feature exhibits strong phase sensitivity (Fig. 2b) and weak dependence on the number of cells, identifying it as a lumped-circuit parametric resonance associated with the discrete implementation rather than with the continuous PTC limit.

The measured broadband amplification profile is qualitatively distinct from conventional resonant amplifier lineshapes. Instead of a Lorentzian resonance with slowly decaying spectral tails, the MBG response forms a broad distributed amplification band inherited from the compact-support instability profile of the ideal infinite PTC. In the finite TMC, losses, parasitics, N-cell implementation, and impedance mismatch smooth the ideal semicircular instability into a continuous asymmetric non-Lorentzian response. The measured spectra are accurately reproduced using a Pearson type IV distribution (shpaw parameter m=0.5), capturing the transformation of the ideal Floquet instability band into a finite experimentally measurable terminal-gain profile. Importantly, the broadband MBG amplification is not a cavity resonance superimposed on the transmission spectrum. Rather, the Floquet instability reorganizes the finite-system response, suppressing temporal Fabry–Pérot oscillations inside the $k$-gap and forcing the spectrum to follow the underlying MBG envelope. The distinct physical origin of the two gain mechanisms is further confirmed by their modulation-phase dependence (Fig. 2b). The broadband MBG amplification remains nearly invariant under modulation-phase shifts, consistent with its origin as a translationally invariant Floquet band-structure effect. In contrast, the narrow mid-gap resonance varies strongly with modulation phase, demonstrating its resonant character and its sensitivity to local interference conditions within the finite lumped circuit. This experimentally separates the robust non-resonant PTC amplification from the additional phase-sensitive parametric instability introduced by the discrete TMC realization.

Further evidence for the MBG origin of the broadband gain is provided by the measured transport dynamics (Fig. 2c). In ideal PTCs, the energy velocity $v_E$ is predicted to go to zero inside the momentum band gap, while the Floquet group velocity becomes divergent near the gap edges [17]. Recent theoretical work showed that the physically meaningful transport quantity in a PTC is the energy velocity rather than the slope of the Floquet dispersion itself, and established that $v_E$ must approach zero at the MBG center even when $d\Omega/dk$ becomes singular [17].

The experimentally extracted group-delay behavior in the finite TMC follows the same qualitative trend. Both measurements and simulations show substantial slowing of the propagating signal inside the broadband gain region, with the minimum velocity occurring near the MBG center where amplification is strongest. Unlike the ideal infinite PTC, however, the measured velocity does not collapse to an exact zero floor. Instead, finite losses, finite modulation duration, incomplete Floquet localization, and residual distributed propagation smooth the singular ideal behavior into a localized minimum. Moreover, because the experimental modulation waveform is trapezoidal, the measured slowing profile becomes asymmetric. This asymmetry follows directly from the asymmetric transport shoulders appearing in the real part of the calculated Floquet dispersion for the trapezoidal modulation (Fig. 1a). Despite these finite-system corrections, the measured velocity profile closely tracks the theoretically predicted PTC energy-velocity behavior, confirming that the observed broadband amplification originates from MBG Floquet physics rather than from resonant enhancement.

Together, these measurements establish that finite microwave TMC systems inherit the defining properties of photonic time crystals: non-resonant broadband amplification, momentum-band-gap transport slowing, and continuous terminal gain arising directly from Floquet instability over a finite frequency interval. At the same time, the experiments reveal how finite-size effects and lumped-circuit inhomogeneities reshape the ideal PTC response into experimentally accessible transmission characteristics.

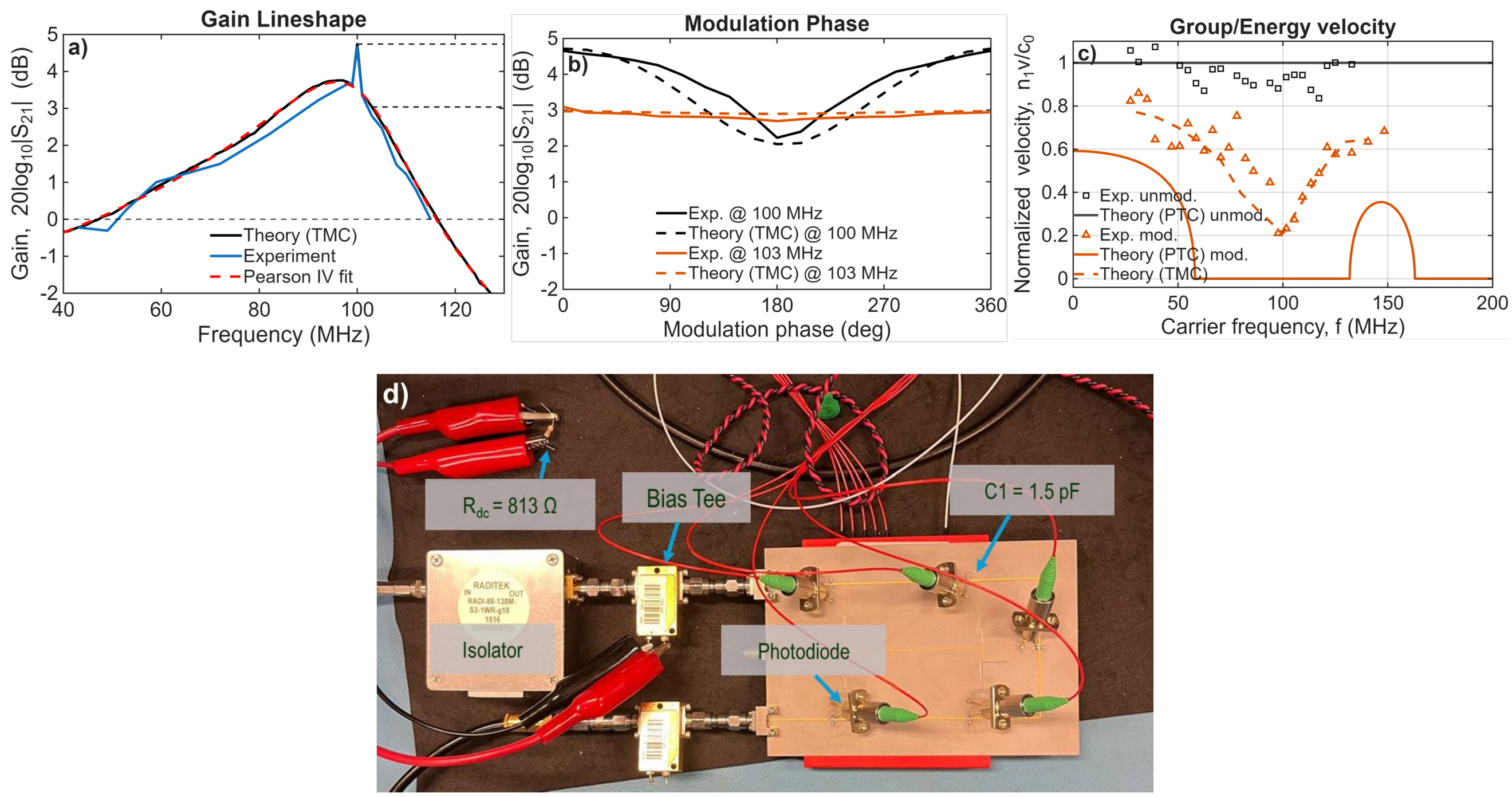

**Figure 2 | Experimental demonstration of PTC amplification in a TMC circuit.** (a) Measured and simulated transmission spectra for a five-cell optically modulated TMC confirm two distinct amplification mechanisms. A broadband gain region emerges around $f_m/2$, consistent with the momentum-band-gap (MBG) predicted by the PTC band structure. Superimposed on this response is a narrow mid-gap parametric resonance associated with the discrete lumped implementation. The broadband MBG amplification follows a continuous asymmetric non-Lorentzian profile accurately reproduced by a Pearson type IV distribution ($m = 0.5$). (b) Modulation-phase dependence separates the two amplification mechanisms. The broadband MBG gain remains nearly invariant under modulation-phase shifts, confirming its MBG origin. In contrast, the narrow mid-gap resonance varies strongly with modulation phase, demonstrating its resonant character and sensitivity to local interference conditions within the finite lumped circuit. (c) Measured and simulated transport behavior inside the gain region further confirms the MBG origin of the amplification. The propagating signal slows substantially inside the broadband gain band, reaching its minimum velocity near the MBG center where amplification is strongest. Although the velocity does not collapse to the strict zero floor predicted for an ideal infinite PTC, the measured profile closely follows the theoretically predicted PTC energy-velocity behavior. [17] (d) Experimental setup, described in the methods section.

## METHODS (Experiment, Figure 2d)

**Device Fabrication.** The PTC circuit was implemented as a periodically loaded microstrip transmission line fabricated on Rogers TC350 Plus substrate (relative permittivity ε ≈ 3.5, loss tangent tanδ ≈ 0.0017 at 10 GHz, thickness 60 mil). Each unit cell consists of a microstrip segment of length $L$ = 57.8 mm, corresponding to approximately λ/16 at the 200 MHz modulation frequency. At the electrical midpoint of each unit cell, the transmission line is loaded with a fixed capacitor ($C_p$ = 1.5 pF) connected in parallel with an optically controlled photodiode.

Both single-cell and five-cell prototypes were fabricated using standard printed circuit board (PCB) manufacturing processes. The structures were terminated with matched $Z_0$ = 50 Ω ports to enable calibrated transmission and gain measurements. DC bias networks, implemented using large-value resistors, were included to provide photodiode biasing while isolating the RF signal path.

**Optical Time Modulation.** Time modulation of the transmission-line impedance was achieved using high-speed InGaAs PIN photodiodes (Thorlabs FGA01FC) operated as optically controlled variable capacitors. The photodiodes were reverse-biased and illuminated using an externally modulated laser source. Optical modulation at 200 MHz was generated by an RF signal generator driving the laser diode and distributed to all unit cells using a 1×5 optical fiber splitter, ensuring synchronized modulation across the structure.

The modulation waveform was a symmetric trapezoid with ~1 ns rise and fall times, preserving a strong fundamental component close to the square-wave limit and thus enabling a large bandgap. Higher harmonics decay more slowly than for sinusoidal modulation, but are not phase-matched as in the quarter-wave square case and therefore leak energy. Under optical illumination, the photodiodes provided deep temporal modulation of the effective capacitance of the transmission line, enabling the formation of a MBG. Detailed photodiode characterization, including capacitance, series resistance, and switching speed as functions of optical power and bias voltage, is provided in the Supplementary Materials.

## METHODS (Theory)

**Analytical and numerical Floquet solutions.** *The infinite PTC and distributed TMC systems* are analyzed using Floquet formalism for temporally periodic media. For square-wave modulation under the temporal quarter-wave condition, closed-form expressions for the Bloch frequency are employed and used as a reference for the gain profile. In general case, the corresponding Hill's equation is solved numerically.

**Temporal transfer-matrix method.** *Finite PTC* structures are analyzed using a temporal transfer-matrix (TMM) formulation, in which each temporal interface is represented by a scattering matrix enforcing field continuity. Cascading $N$ modulation periods yields the overall response, from which transmission and gain are extracted. This approach accounts for finite temporal truncation and temporal boundary reflections.

**Band-structure solver for TMC.** *For the lumped periodically time-modulated TMC systems*, the Floquet band structure is obtained using a transfer-matrix formulation over one modulation period with periodic boundary conditions imposed in time. The modulation cycle is discretized into a sequence of temporal states, whose transfer matrices are cascaded to construct the total evolution matrix. The Floquet harmonic expansion is truncated to a finite number of harmonics to obtain a finite-dimensional eigenvalue problem, whose eigenvalues determine the complex Bloch frequency. To relate the lumped transmission-line realization to the corresponding distributed photonic time-crystal description, the dispersion is mapped through the long-wavelength approximation $\sin(kd/2) \sim kd/2$. The lumped TMC model exhibits a narrow mid-gap parametric resonance near $f_m/2$, originating from the discrete circuit implementation.

**FDTD simulation of the transmission line.** A one-dimensional FDTD scheme is implemented for the time-modulated transmission line, solving the telegrapher's equations with time-dependent capacitance. The system is excited at a fixed carrier frequency, and steady-state fields are extracted after transients decay. The transmitted signal is evaluated at the output port, from which the transmission and gain spectra are obtained and compared with band-structure predictions.

## Conclusion

We have experimentally demonstrated that the exponential Floquet instability associated with a photonic time-crystal momentum band gap can survive the finite-size, lossy, and strongly inhomogeneous conditions of a practical microwave platform and emerge as stable positive terminal gain over a continuous broadband frequency interval. Using an optically modulated time-modulated-capacitor (TMC) transmission-line system, we observed broadband non-resonant amplification consistent with momentum-band-gap physics, together with transport slowing and phase-invariant gain behavior inherited directly from the underlying PTC band structure.

The experiments further reveal how finite practical implementations reshape the ideal PTC response. In the infinite homogeneous limit, the instability follows a compact-support semicircular gain envelope. The finite TMC implementation transforms this ideal gain shape into a continuous asymmetric non-Lorentzian terminal-gain profile accurately described by a Pearson type IV distribution. At the same time, the discrete implementation introduces an additional narrow phase-sensitive parametric resonance at the MBG center, distinct from the broadband Floquet amplification itself. These observations establish a direct physical connection between ideal photonic time-crystal theory and experimentally measurable transmission gain in finite propagating systems.

The demonstrated TMC architecture therefore serves primarily as a proof-of-principle experimental realization linking ideal PTC Floquet physics to practical microwave circuitry. Having established this connection, future work can now focus on the engineering limits of such systems, including scaling toward GHz modulation rates, increasing the achievable modulation depth, improving distributed homogeneity, and optimizing independently for bandwidth or peak gain. Preliminary studies indicate that substantially larger gain values may be achievable in optimized architectures while preserving the distributed non-resonant MBG amplification mechanism demonstrated here.

More broadly, these results pave the way toward a new class of Floquet-driven RF and microwave amplifiers operating through distributed temporal-instability physics rather than conventional resonant or transistor-based amplification mechanisms. By establishing experimentally that momentum-band-gap amplification can produce stable broadband terminal gain in a finite system, this work provides an experimentally accessible foundation for future time-crystal-based active photonic and microwave devices.

**Contribution**. TRJ developed and characterized the experimental platform, including systematic high-precision measurements calibrated against device-level simulations, enabling close quantitative agreement and direct comparison with theory. MS provided foundational expertise and guidance on photonic time-crystal physics, including conceptual development throughout the work. LJP supported the project through theoretical/numerical analysis, interpretation, and manuscript drafting. AVK and DP contributed through scientific discussions, supervision, and funding acquisition. DP led and supported the laboratory and experimental research team. All authors contributed to writing.

## References

[1] Lyubarov, M.; Lumer, Y.; Dikopoltsev, A.; Lustig, E.; Sharabi, Y.; Segev, M. *Amplified emission and lasing in photonic time crystals.* Science **2022**, *377*, 425–428.

[2] Lustig, E.; Segal, O.; Saha, S.; Fruhling, C.; Shalaev, V. M.; Boltasseva, A.; Segev, M. *Photonic time-crystals – fundamental concepts.* Opt. Express **2023**, *31*, 9165–9170.

[3] Galiffi, E.; Tirole, R.; Yin, S.; Li, H.; Vezzoli, S.; Huidobro, P. A.; Silveirinha, M. G.; Sapienza, R.; Alù, A.; Pendry, J. B. *Photonics of time-varying media.* Adv. Photonics **2022**, *4*, 014002.

[4] Engheta, N. *Four-dimensional optics using time-varying metamaterials.* Science **2023**, *379*, 1190–1191.

[5] Pacheco-Peña, V.; Solís, D. M.; Engheta, N. *Time-varying electromagnetic media: opinion.* Opt. Mater. Express **2022**, *12*, 3829–3840.

[6] Caloz, C.; Deck-Léger, Z.-L. *Spacetime metamaterials—Part II: Theory and applications.* IEEE Trans. Antennas Propag. **2020**, *68*, 1583–1612.

[7] Solís, D. M.; Kastner, R.; Engheta, N. *Time-varying materials in the presence of dispersion: plane-wave propagation in a Lorentzian medium with temporal discontinuity.* Photonics Res. **2021**, *9*, 1842–1853.

[8] Galiffi, E.; Martínez Solís, D.; Yin, S.; Engheta, N.; Alù, A. *Electrodynamics of photonic temporal interfaces.* Light Sci. Appl. **2025**, *14*, 338.

[9] Li, Z.; Ma, X.; Bahrami, A.; Deck-Léger, Z.-L.; Caloz, C. *Space-time Fresnel prism.* Phys. Rev. Applied **2023**, *20*, 054029.

[10] Pacheco-Peña, V.; Engheta, N. *Spatiotemporal cascading of dielectric waveguides.* Opt. Mater. Express **2024**, *14*, 1062–1073.

[11] Rizza, C.; Castaldi, G.; Galdi, V. *Nonlocal effects in temporal metamaterials.* Nanophotonics **2022**, *11*, 1285–1295.

[12] Sharabi, Y.; Lustig, E.; Segev, M. *Disordered photonic time crystals.* Phys. Rev. Lett. **2021**, *126*, 163902.

[13] Galiffi, E.; Harwood, A. C.; Vezzoli, S.; Tirole, R.; Alù, A.; Sapienza, R. *Optical coherent perfect absorption and amplification in a time-varying medium.* Nat. Photonics **2026**, *20*, 163–169.

[14] Solís, D. M.; Engheta, N. *Functional analysis of the polarization response in linear time-varying media: A generalization of the Kramers–Kronig relations.* Phys. Rev. B **2021**, *103*, 144303.

[15] Globosits, D.; Hüpfl, J.; Rotter, S. *A photonic Floquet scattering matrix for wavefront-shaping in time-periodic media.* arXiv **2024**, arXiv:2403.19311.

[16] Garg, P.; Lamprianidis, A. G.; Rahman, S.; Stefanou, N.; Almpanis, E.; Papanikolaou, N.; Verfürth, B.; Rockstuhl, C. *Two-step homogenization of spatiotemporal metasurfaces using an eigenmode-based approach.* Opt. Mater. Express **2024**, *14*, 549–563.

[17] Lee, K.; Kim, Y.; Kim, K. W.; Min, B. *Energy transport velocity in photonic time crystals.* arXiv **2026**, arXiv:2602.03453.

[18] Moussa, H.; Xu, G.; Yin, S.; Galiffi, E.; Ra'di, Y.; Alù, A. *Observation of temporal reflection and broadband frequency translation at photonic time interfaces.* Nat. Phys. **2023**, *19*, 863–868.

[19] Jones, T. R.; Kildishev, A. V.; Segev, M.; Peroulis, D. *Time-reflection of microwaves by a fast optically-controlled time-boundary.* Nat. Commun. **2024**, *15*, 6786.

[20] Wang, X.; Mirmoosa, M. S.; Asadchy, V. S.; Rockstuhl, C.; Fan, S.; Tretyakov, S. A. *Metasurface-based realization of photonic time crystals.* Sci. Adv. **2023**, *9*, eadg7541.

[21] Xiong, J.; Zhang, X.; Duan, L.; Wang, J.; Long, Y.; Hou, H.; Yu, L.; Zou, L.; Zhang, B. *Observation of wave amplification and temporal topological state in a non-synthetic photonic time crystal.* Nat. Commun. **2025**, *16*, 11182.

[22] Liu, T.; Ou, J.-Y.; MacDonald, K. F.; Zheludev, N. I. *Photonic metamaterial analogue of a continuous time crystal.* Nat. Phys. **2023**, *19*, 986–992.